\documentclass[twocolumn]{aastex63}

\usepackage{savesym}
\usepackage{newtxtext,newtxmath}
\usepackage{amsmath, amssymb, graphics, bm, graphicx}
\allowdisplaybreaks

\newcommand{\mathsym}[1]{{}}
\newcommand{\unicode}[1]{{}}

\newcommand{\ag}{\alpha}
\newcommand{\bg}{\beta}

\newcommand{\dg}{\delta}
\newcommand{\Dg}{\Delta}

\newcommand{\om}{\omega}

\newcommand{\der}{{\rm d}}

\newcommand{\vphi}{\varphi}
\newcommand{\eps}{\epsilon}

\newcommand{\hI}{{\bm {\hat I}}}
\renewcommand{\hom}{{\bm {\hat \omega}}}

\newcommand{\hn}{{\bm {\hat n}}}

\newcommand{\hp}{{\bm {\hat p}}}
\newcommand{\bom}{{\bm \omega}}
\newcommand{\bL}{{\bm L}}

\newcommand{\bIeff}{{\bf I}_{\rm eff}}

\newcommand{\bcdot}{{\bm \cdot}}
\newcommand{\btimes}{{\bm \times}}

\newcommand{\Pprec}{P_{\rm prec}}
\newcommand{\epseff}{\epsilon_{\rm eff}}
\newcommand{\epsp}{\epsilon_{\rm p}}
\newcommand{\epspar}{\epsilon_\parallel}
\newcommand{\epsdg}{\epsilon_\delta}

\newcommand{\Bp}{B_{\rm p}}
\newcommand{\Bs}{B_\star}

\newcommand{\bR}{\bar R_6}
\newcommand{\bM}{\bar M_{1.4}}

\newcommand{\tobs}{t_{\rm obs}}
\newcommand{\tsd}{t_{\rm sd}}

\newcommand{\Msun}{{\rm M}_\odot}
\newcommand{\cm}{{\rm cm}}
\newcommand{\G}{{\rm G}}
\newcommand{\s}{{\rm s}}
\newcommand{\days}{{\rm days}}
\newcommand{\years}{{\rm years}}

\newcommand{\be}{\begin{equation}}
\newcommand{\ee}{\end{equation}}

\shorttitle{Periodic FRBs with NS Precession}
\shortauthors{Zanazzi \& Lai}
\graphicspath{{./}{figures/}}

\begin{document}

\title{Periodic Fast Radio Bursts with Neutron Star Free Precession}

\correspondingauthor{J. J. Zanazzi}
\email{jzanazzi@cita.utoronto.ca}

\author{J. J. Zanazzi}
\affiliation{Canadian Institute for Theoretical Astrophysics,
University of Toronto,
60 St. George Street,
Toronto, Ontario, M5S 1A7, Canada}

\author{Dong Lai}
\affiliation{Department of Astronomy,
Center for Astrophysics and Planetary Science,
Cornell University,
Ithaca, NY 14853, USA}
\affiliation{Department of Astronomy and Miller Institute for Basic Research In Science, UC Berkeley, Berkeley, CA 94720, USA}

\begin{abstract}
The CHIME/FRB collaboration recently reported the detection of a 16
day periodicity in the arrival times of radio bursts from FRB
180916.J0158+65. We study the possibility that the observed
periodicity arises from free precession of a magnetized neutron star,
and put constraints on different components of the star's magnetic
fields.  Using a simple geometric model, where radio bursts are emitted from a rotating neutron star magnetosphere, we show 
that the emission pattern as a function of time can match that observed from FRB
180916.J0158+65.
\end{abstract}

\keywords{
radiation mechanisms: general -- polarization -- stars: neutron -- stars: magnetars
}


\section{Introduction}

Fast radio bursts (FRBs) are extragalactic milli-second radio transients,
and their origin is mysterious
\citep{Katz(2018)rev,Petroff(2019),CordesChatterjee(2019)}.  
An increasing number of FRBs have been found to repeat
\citep{CHIME(2019)}.
So far, no sign of periodicity has been detected in any FRBs (such as FRB 121101, \citealt{Zhang(2018),Katz(2018)}). Recently, the Canadian Hydrogen Intensity Mapping Experiment Fast Radio Burst
Project (CHIME/FRB) team reported the detection of periodicity
from a repeating FRB 180916.J0158+65 (hereafter FRB 180916; \citealt{CHIME(2020)}):
The 28 bursts recorded by CHIME in the 410 days timespan (from 9/2018
to 10/2019) exhibit a period of $16.35\pm 0.18$ days in arrival times,
and cluster in a $\sim$4-day phase window.  This finding, if confirmed by
future observations and found to be generic for many FRBs, would
provide a significant clue to the nature of these objects.

The CHIME discovery paper already discussed several possible origins
for the periodicity, including pulsars in binaries and isolated
precessing neutron stars.  
In this paper we examine the latter
possibility and the implication for the central engine of FRBs (see also \citealt{Levin(2020)}).

Neutron star (NS) precession has long been studied in the literature.
It was recognized early on that superfluid vortex pinning in the NS
crust suppresses free precession \citep{Shaham(1977)}.
Revised superfluid properties or the absence of superfluidity may still allow
precession to occur \citep{LinkEpstein(1997),Sedrakian(1999),Akgun(2006),GoglichidzeBarsukov(2019)}. 
Some observed long-term variabilities of radio pulsar emission
\citep{Kramer(2006),Weisberg(2010),Lyne(2013)}
may be attributed to free precession \citep{ZanazziLai(2015),Arzamasskiy(2015)}. 
Free precession
could also influence the x-ray variability and spindown of
magnetars in the Galaxy \citep{Melatos(1999)}.

This work investigates if NS free precession can explain the periodicity of FRB 180916.  In Section~\ref{sec:PeriodicFRB_NSPrec} we constrain NS magnetic fields from the observed period, and calculate the emission pattern from a simple geometrical FRB model.  
We discuss the effect of precession on linear polarization in Section~\ref{sec:Discuss}
and conclude in Section~\ref{sec:Conc}.

\section{Periodic FRBs from NS Precession}
\label{sec:PeriodicFRB_NSPrec}


\subsection{Free Precession of NS}
\label{sec:NS_Prec}

\begin{figure}
	\includegraphics[width=\columnwidth]{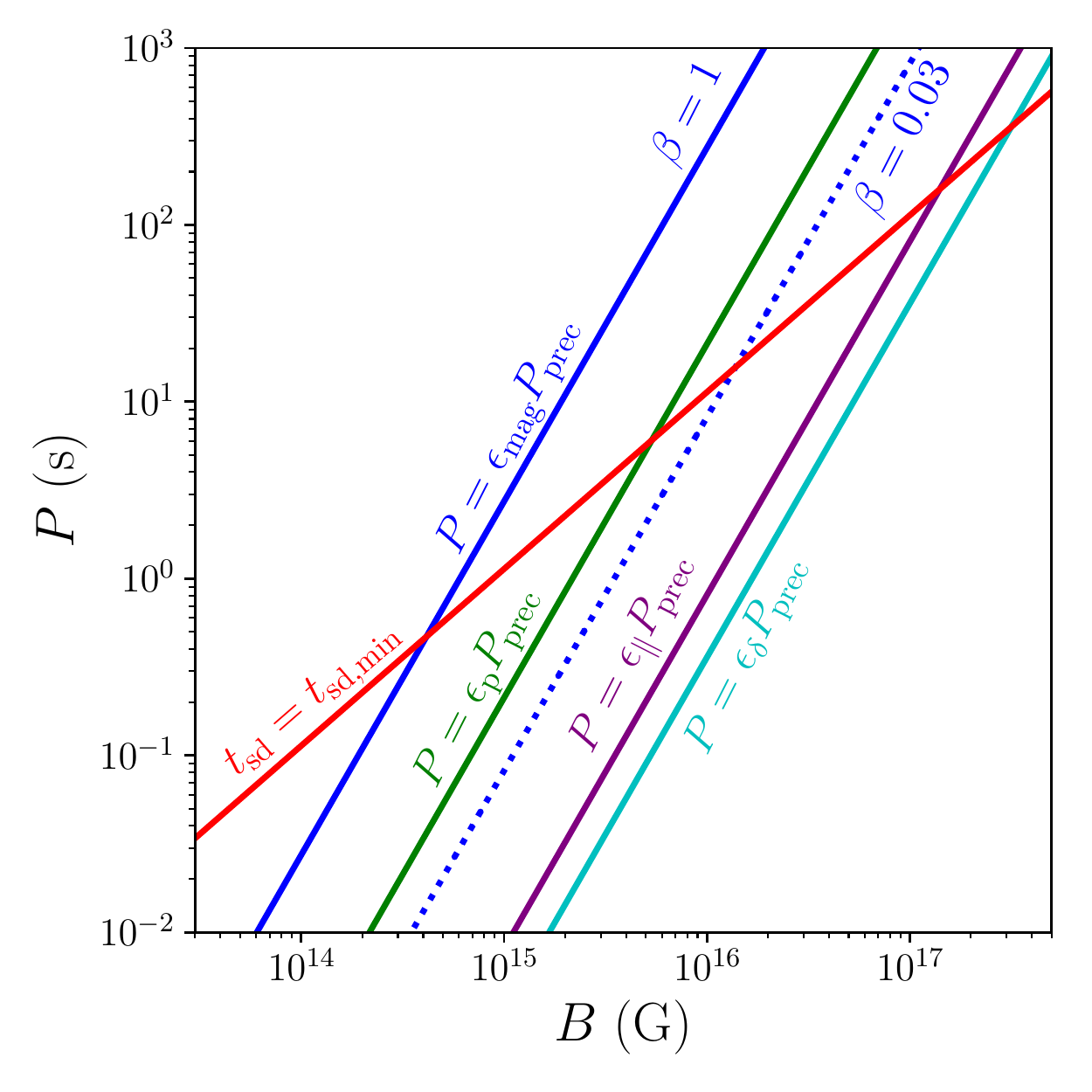}
    \caption{
    NS spin period $P$ and magnetic field strength $B$ which lead to a spin precession period $\Pprec
    =16.35$~days (eq.~\ref{eq:Pprec}), with the effective ellipticity $\epseff = \eps_{\rm mag}$ (blue; eq.~\ref{eq:eps_mag} with $B = \Bs$), $\epseff = \epsp$ (green; eq.~\ref{eq:eps_p} with $B = \Bp$), $\epseff = \epspar$ (purple; eq.~\ref{eq:eps_quad} with $B = B_\parallel$), and $\epseff = \epsdg$ (cyan; eq.~\ref{eq:eps_quad} with $B = B_\dg$).  The red line displays the $P$ and $B = \Bp$ values when the duration over which FRB 180916 was observed ($\tobs = 410$~days) equals the NS spin-down time $\tsd$ (eq.~\ref{eq:tsd}).  Here, $\cos\theta = 1$, with $\bg = 1$ (solid) and $\bg = 0.03$ (dotted).
    }
    \label{fig:P_B_constraints}
\end{figure}

\begin{figure}
	\includegraphics[width=\columnwidth]{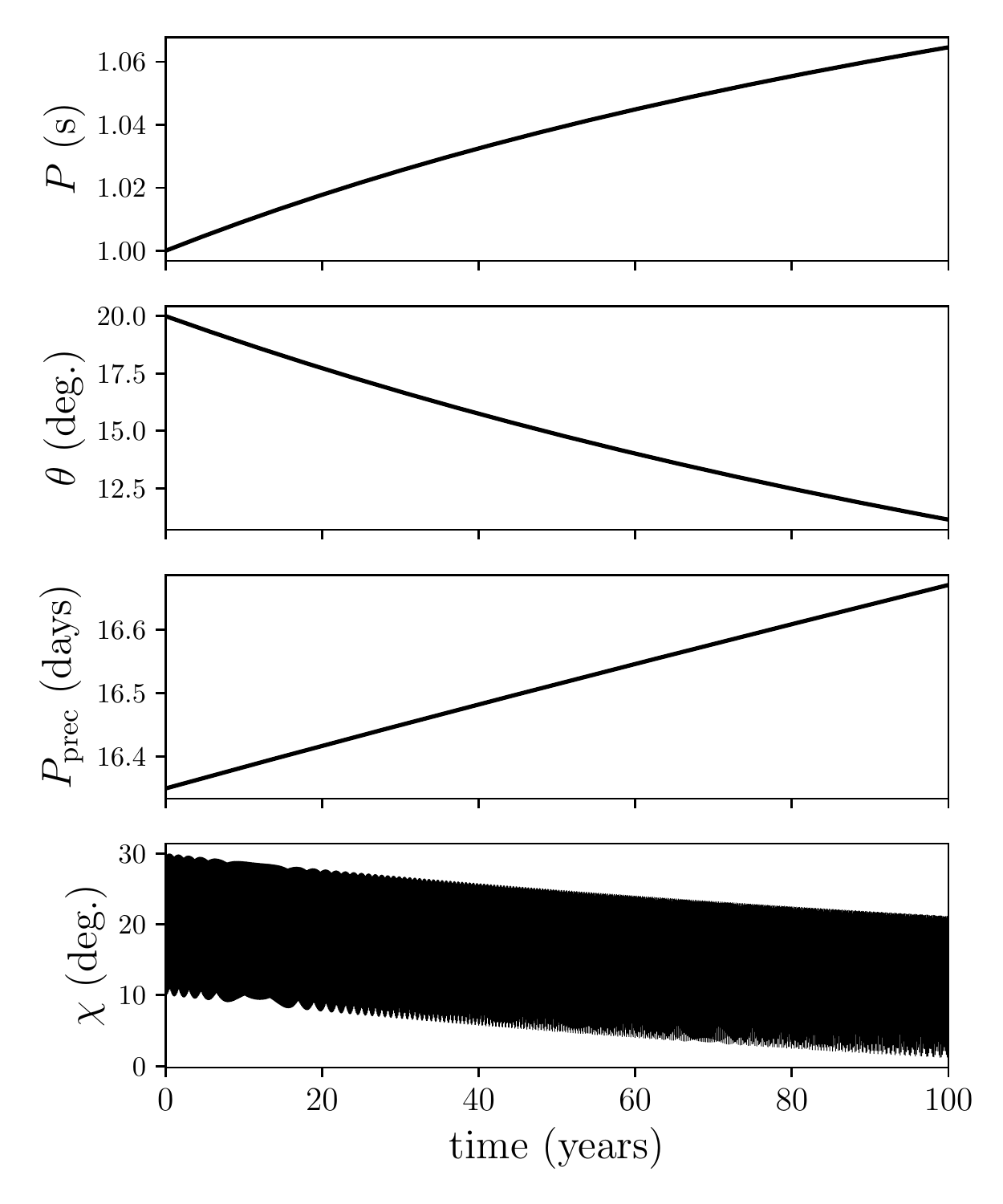}
    \caption{
    Time evolution of the NS spin period $P$ (first panel), precession angle $\theta$ (second panel; see Fig.~\ref{fig:NSPrecDiag}), precession period $\Pprec$ (third panel; eq.~\ref{eq:Pprec}), and magnetic inclination angle $\chi$ (last panel; angle between $\hp$ and $\hom$).  We evolve the NS spin frequency $\om = 2\pi/P$ and $\theta$ using equations~\eqref{eq:domdt_sd} and~\eqref{eq:dthetadt_sd}, with $\chi$ evaluated using equation~\eqref{eq:hom(t)}.  Here, $M = 1.4 \, \Msun$, $R = 10^6 \, \cm$, $\Bp = 10^{15} \, \G$, $P(0) = 1 \, \s$, $\theta(0) = 20^\circ$, $\psi = 10^\circ$, and $\epseff = 7.53 \times 10^{-7}$.
    }
    \label{fig:SpinDown}
\end{figure}

Consider a NS with mass $M$, radius $R$, dipolar magnetic field of strenth $B_{\rm p}$, dipole moment $p = \frac{1}{2} \Bp R^3$ and axis $\hp$, spin period $P$, and spin frequency $\om = 2\pi/P$.  The NS could also have complex quadrupole field and internal fields (see below). For simplicity, we assume the NS is homogeneous, with a constant density $\rho = 3 M/(4\pi R^3)$ and 
moment of inertia $I = \frac{2}{5} M R^2$.  We define $\bR = R/(10^6 \, {\rm cm})$ and $\bM = M/(1.4 \, \Msun)$.

In the frame rotating with the NS, the equations of motion describing the evolution of the NS spin vector $\bom = \om \hom$ is \citep{ZanazziLai(2015)}
\be
\frac{\der \bL_{\rm eff}}{\der t} + \bom \btimes \bL_{\rm eff} = 0,
\label{eq:Euler}
\ee
where $\bL_{\rm eff} = \bIeff \bcdot \bom$ is the effective angular momentum of the NS. The effective moment of inertia tensor $\bIeff$ takes account of the non-sphericity of the NS due to rotation and internal magnetic fields, as well as the inertia 
from the near-zone fields corotating with the NS \citep{DavisGoldstein(1970),Goldreich(1970),ZanazziLai(2015)}. Take $I_i$ to be the eigenvalues of $\bIeff$ (effective principal moments of inertia), with $\hI_i$ their associated unit eigenvectors (effective principal axes).  For simplicity, we assume $I_1 = I_2$, but $\epseff = (I_3 - I_1)/I_1 \ne 0$ (biaxial NS; we assume $|\epseff| \ll 1$ throughout).  Then equation~\eqref{eq:Euler} has the solution \citep{LandauLifshitz(1969), Goldreich(1970)}
\be
\hom = \sin \theta \cos \vphi_\om \hI_1 + \sin \theta \sin \vphi_\om \hI_2 + \cos \theta \hI_3,
\label{eq:hom(t)}
\ee
where
\be
\vphi_\om(t) = (\epseff \om \cos\theta) t  + \vphi_{\om 0}
\label{eq:vphi_om(t)}
\ee
is the precession phase of $\hom$ around $\hI_3$, with $\vphi_{\om 0}=\vphi_\om(0)$, while $\theta$ is the angle between $\hom$ and $\hI_3$ ($\cos \theta = \hom \bcdot \hI_3$).  Notice $\om$ and $\theta$ are constants of motion for equation~\eqref{eq:Euler}.  
The NS precession period $\Pprec$ is then
\be
\Pprec = \frac{P}{\epseff \cos \theta}.
\label{eq:Pprec}
\ee

When $I_1 \ne I_2$, equation~\eqref{eq:Euler} can be solved with qualitatively similar dynamics, except the magnitude of $\om$ oscillates and $\theta$ nutates with time (\citealt{LandauLifshitz(1969),Melatos(1999),ZanazziLai(2015)}, see also \citealt{Levin(2020)} for why triaxiality does not make the precession rate vary for a nearly-spherical rotator).  

We postulate that the observed 16.35 day period seen in FRB 180916 is the precession period $\Pprec$. This constrains 
$\epseff$ to be
\be
\epseff = \frac{7.1 \times 10^{-7}}{\cos \theta} \left( \frac{P}{1 \, \s} \right) \left( \frac{16.35 \, \days}{\Pprec} \right).
\label{eq:epseff_obs}
\ee
There are several contributions to the non-sphericity parameter $\epseff$.  Two primary sources are intrinsic to the NS. The first arises from the internal magnetic field of strength $B_\star$, leading to a deformation of order
\begin{align}
&\eps_{\rm mag} = \bg \frac{R^4 B_\star^2}{G M^2} = 1.9 \times 10^{-6} \bg \left( \frac{B_\star}{10^{15} \, \G} \right)^2 \frac{\bR^4}{\bM^2},
\label{eq:eps_mag}
\end{align}
where $\bg$ is a constant satisfying $|\beta|\ll 1$ (either $\bg > 0$ or $\bg < 0$), with a value which depends on the magnetic field's topology \citep{Mastrano(2013)}; a complex internal field can yield
$|\beta|\ll 1$.  The second deformation source is an elastic crust which has a rotational bulge with principal axis $\hI_3$ misaligned with $\hom$, formed when the crust crystallized at a higher rotational frequency \citep[e.g.][]{Goldreich(1970),Cutler(2003)}.  Assuming the NS has a uniform shear modulus $\mu$, the deformation from elasticity is of order (assuming $19 \mu \ll 2 \rho g R$, where $g = G M/R^2$; \citealt{MunkMacDonald(1975)})
\begin{align}
&\eps_{\rm elast} \simeq \ \left( \frac{19 \mu}{2 \rho g R} \right) \left( \frac{15 \om^2}{16 \pi G \rho} \right)
\nonumber \\
&= 2.0 \times 10^{-11} \left( \frac{\mu}{10^{30} \, {\rm dynes}/{\rm cm}^2} \right) \left( \frac{1 \, \s}{P} \right)^2 \frac{\bR^7}{\bM^3}.
\label{eq:eps_elast}
\end{align}
Although the frozen-in rotational deformation could be larger when the NS was born rotating fast, the NS's spin-down causes the crust to experience stress and break before the NS has slowed to near its present rotation rate \citep[e.g.][]{BaymPines(1971),Cutler(2003)}.  Since the NS may have experienced many crust-quakes over its lifetime, it is reasonable to assume that the frozen-in rotational deformation is of order the present rotational deformation.

In addition to the intrinsic deformations $\eps_{\rm mag}$ and $\eps_{\rm elast}$, 
the near-zone fields corotating with the NS induces a precessional torque
\citep{Goldreich(1970)}, and this effect can be incorporated into the effective deformation parameter
\citep{ZanazziLai(2015)}.
The dipole field gives \citep{Melatos(1999),Melatos(2000),ZanazziLai(2015)}
\be
\eps_{\rm p} = \frac{3 \Bp^2 R^5}{20 I c^2} = 1.5 \times 10^{-7} \left( \frac{\Bp}{10^{15} \, \G} \right)^2 \frac{\bR^3}{\bM}.
\label{eq:eps_p}
\ee
Equation~\eqref{eq:eps_p} takes into account the inertia of the field exterior to the NS in vacuum; including the inertia of the field inside the NS \citep{BeskinZheltoukhov(2014)}, or 
the effect of magnetosphere plasma \citep{Arzamasskiy(2015)}, modifies equation~\eqref{eq:eps_p} by factors of order unity.  Similarly, a quadrupolar magnetic field with strengths specified by $B_\parallel$ and $B_\dg$ leads to deformations of order 
\be
\epspar = \frac{4}{105} \left( \frac{B_\parallel}{\Bp} \right)^2 \epsp, \hspace{5mm} \epsdg = \frac{16}{945}\left( \frac{B_\dg}{\Bp} \right)^2 \epsp,
\label{eq:eps_quad}
\ee
see \cite{ZanazziLai(2015)} for details and definitions of $B_\parallel$ and $B_\dg$.  For magnetic field strengths ($B \sim 10^{15} \, \G$) and spin periods ($P \sim 1 \, \s$) typical of magnetars, we see $\eps_{\rm mag}$, $\epsp$, $\epspar$, and $\epsdg$ are all feasible ways to effectively deform the NS to give a spin precession period $\Pprec = 16.35 \ \days$, but elastic deformation $\eps_{\rm elast}$ requires $P \sim 1 \, {\rm ms}$ to get $\eps_{\rm elast} \sim \epseff$.  Since this is much shorter than a typical magnetar $P$ value, we will not consider $\eps_{\rm elast}$ for the remainder of this work.

In equation~\eqref{eq:Euler}, we have neglected the radiative torque, which works to spin down the NS  and secularly align $\hom$ with $\hp$.  This is valid as long as the shift in the NS precession  phase due to spin-down over the course of the observation ($\Dg \vphi_\om \sim 2\pi \tobs^2/[\Pprec \tsd]$) is less than unity, where
\be
t_{\rm sd} = \frac{3 c^3 I}{2 p^2 \om^2}= 145 \left( \frac{P}{1 \, \s} \right)^2 \left( \frac{10^{15} \, \G}{\Bp} \right)^2 \frac{\bM}{\bR^4} \, \years
\label{eq:tsd}
\ee
is the spin-down time for the NS, and $\tobs = 410$~days is the length of time which FRB 180916 was observed. Requiring $\Dg \vphi_\om \lesssim 1$ gives the constraint $\tsd \gtrsim 2\pi \tobs^2/\Pprec \equiv t_{\rm sd,min}$ for our precession model.\footnote{Note that the precession period is always less than $t_{\rm sd}$ by a factor
$\lesssim \omega R/c$, since $\epseff \gtrsim \epsp$, and thus $\Pprec \lesssim P/\epsp \sim (\omega R/c) \tsd$.}.

Figure~\ref{fig:P_B_constraints} depicts the constraints on the NS spin period $P$ and the strengths of various magnetic field components (internal, dipole and quadrupole) in order for magnetic deformations (both intrinsic and effective) to produce $\Pprec = 16.35 \ \days$.  For spin periods $P \sim 0.1-10 \ \s$,  a range of magnetic field values
($B \sim 10^{14} - 10^{17}$) are required, depending on which deformation mechanism dominates
$\epseff$. Figure~\ref{fig:P_B_constraints} also shows the condition $\tsd = t_{\rm sd,min}$. For the NS to stably precess over the observed duration of FRB 180916 ($\tobs$) with a given poloidal field $\Bp$, the $P$ value must lie somewhat above the red line.

Over timescales comperable to $\tsd$, the NS spin frequency $\om$ and precession angle $\theta$ evolve due to the radiative torque.  When $\tsd \gg \Pprec$, the evolutionary equations for $\om$ and $\theta$ are \citep{Goldreich(1970),ZanazziLai(2015)}
\begin{align}
    \frac{\der \om}{\der t} &= - \frac{\om}{\tsd} \left[ \sin^2 \psi + \sin^2 \theta \left( 1 - \frac{3}{2} \sin^2 \psi \right) \right],
    \label{eq:domdt_sd} \\
    \frac{\der \theta}{\der t} &= - \frac{1}{\tsd} \cos \theta \sin \theta \left( 1 - \frac{3}{2} \sin^2 \psi \right).
    \label{eq:dthetadt_sd}
\end{align}
Figure~\ref{fig:SpinDown} depicts an example of the evolution of 
$P$, $\theta$, $\Pprec$, and $\chi$ (the magnetic inclination angle, or angle between $\hom$ and the dipole axis $\hp$)
over timescales comparable to $\tsd$. For the example given, 
$P$ and $\Pprec$ increase, while $\theta$ and $\chi$ decrease, with time.

\subsection{Model for FRB Emission from Precessing NS}
\label{sec:FRBEmit}

\begin{figure}
	\includegraphics[width=\columnwidth]{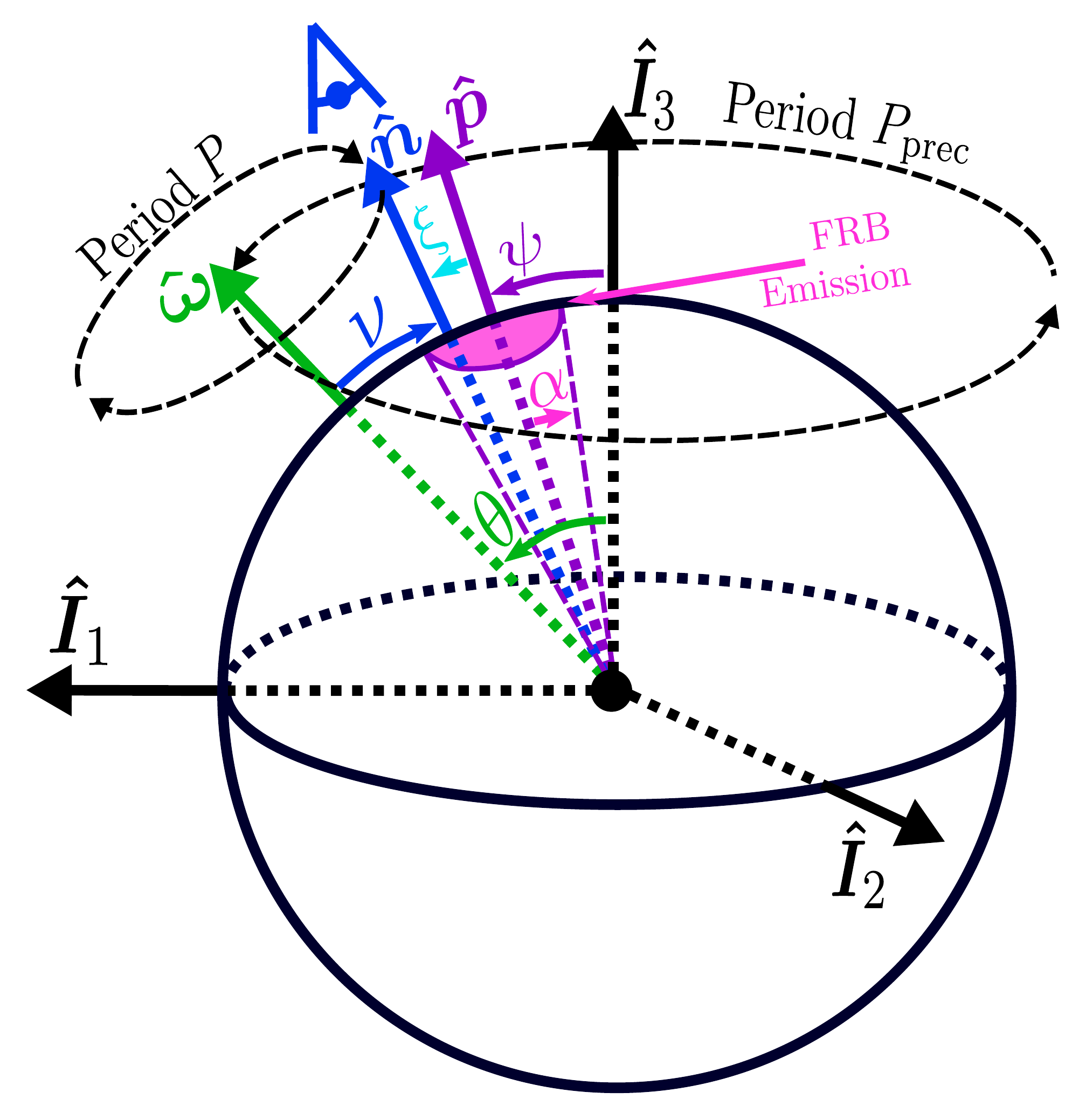}
    \caption{
    Geometric model of FRB emission from a precessing NS, depicted in the rotating frame (body frame) of the NS. See text for details.
    }
    \label{fig:NSPrecDiag}
\end{figure}

\begin{figure*}
    \centering
	\includegraphics[scale=0.57]{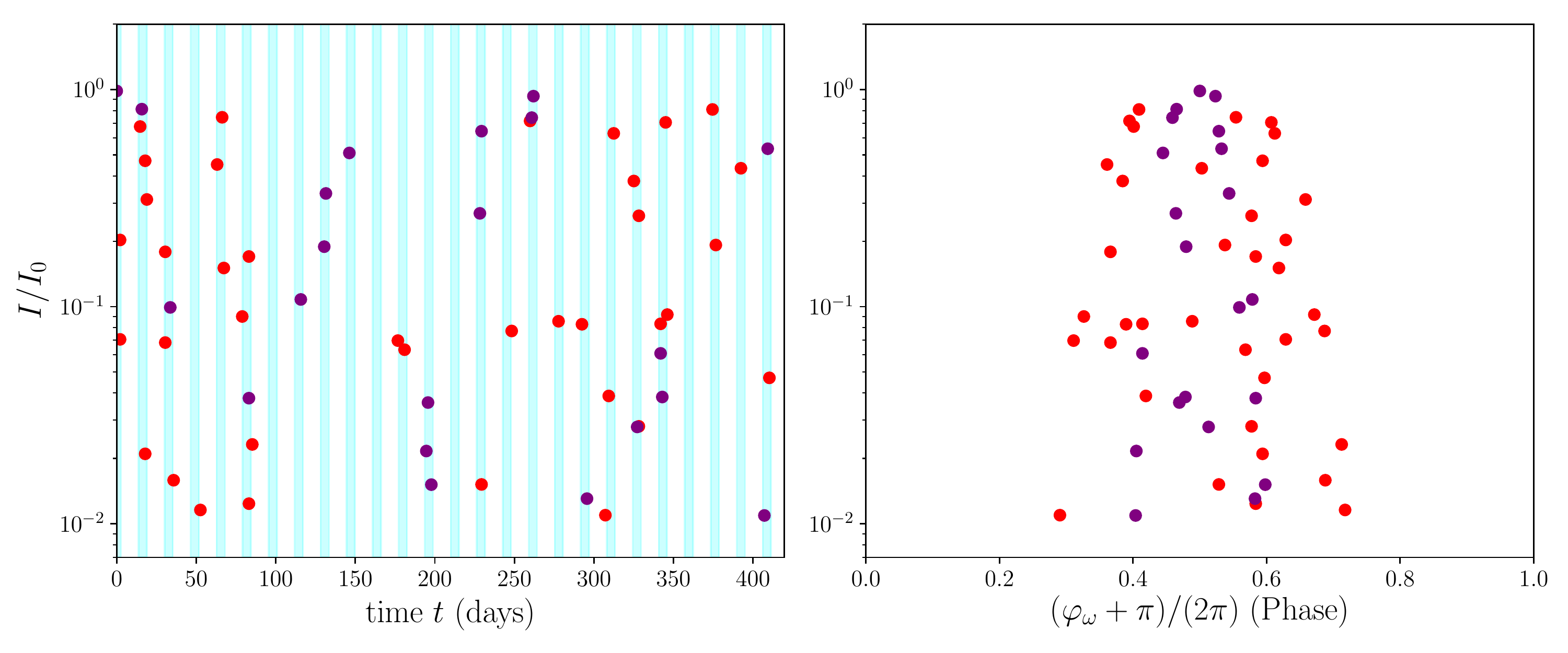}
    \caption{
    FRB emission $I$ over time $t$ (left panel) and precession phase $\vphi_\om$ (eq.~\ref{eq:vphi_om(t)}; right panel) for our precessing NS model over the observed duration of FRB 180916 ($\tobs = 410$~days), with the spin precession frequency set to FRB 180916's period ($\Pprec = 16.35 \ \days$).  Dots show $I$ evaluated at times $t_i$, while light cyan vertical bands denote $\pm 2.6$ day intervals around multiples of $\Pprec$, which are the epochs where emission from FRB 180916 was detected.  At every time $t = t_i$, the rotational phase $\vphi_n$ (eq.~\ref{eq:vphi_n}) is drawn randomly from the uniform interval $[0,2\pi]$.  The model parameters are $\psi = 3^\circ$ (red), $\psi = 7^\circ$ (purple), with $\theta = 10^\circ$, $\nu = \theta-\psi$, $\ag = 1^\circ$, $N = 400$, and $\vphi_{\om 0} = 0$.  Our model assumes $I_0$ is fixed: all $I$ variations with $t$ are due to observing the FRB off-center ($\hn \ne \hp$, see eq.~\ref{eq:I}).
    }
    \label{fig:NSEmit}
\end{figure*}

\begin{figure*}
    \centering
	\includegraphics[scale=0.57]{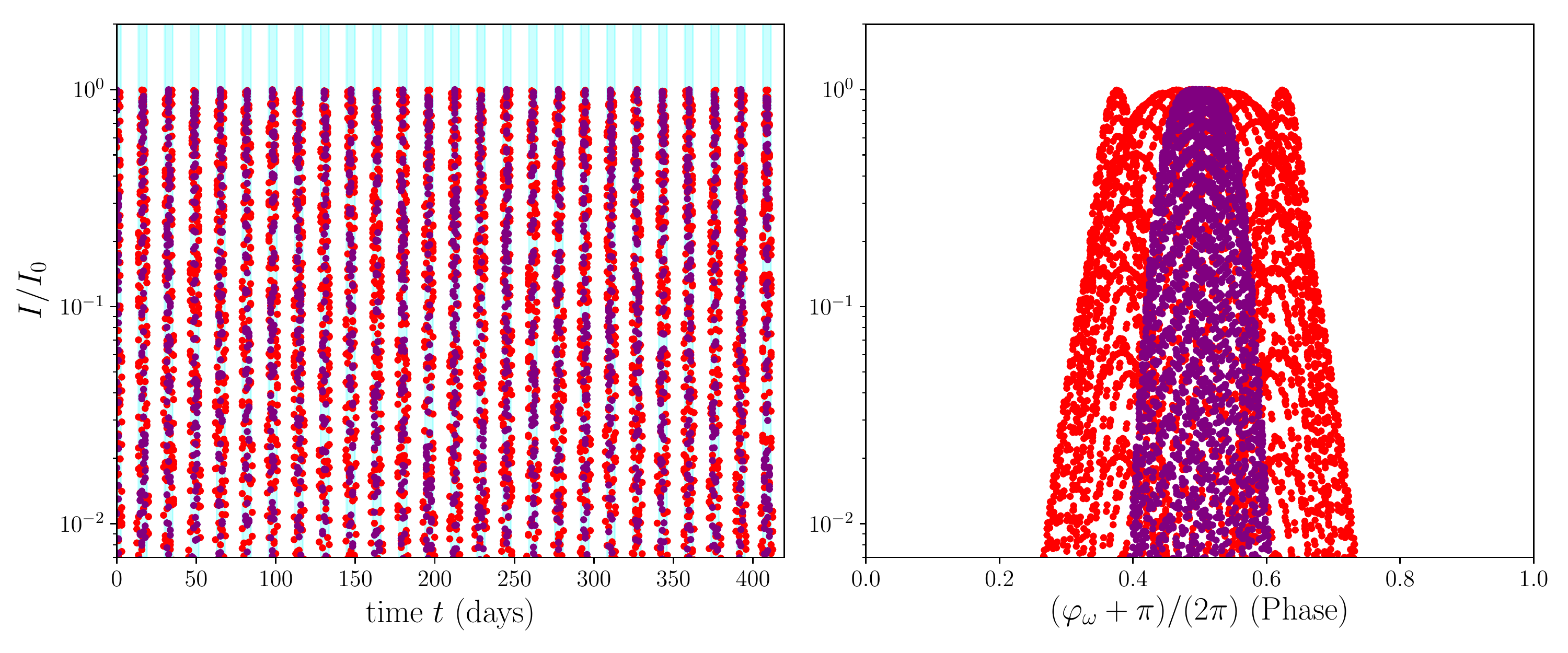}
    \caption{
    Same as Figure~\ref{fig:NSEmit}, except at every time $t = t_i$, we pick $N_n=100$ linearly-spaced values for the precession phase $\vphi_n$ (eq.~\ref{eq:vphi_om(t)}) spanning the interval $[0,2\pi]$.
    }
    \label{fig:NSEmitPhase}
\end{figure*}

The central engine of FRBs and the radiation mechanism are uncertain. Given the millisecond timescale of the radio bursts, it is natural that most models associate FRB emissions to magnetized
neutron stars \citep[e.g.][]{Lyubarsky(2014),CordesWasserman(2016),Katz(2016),Beloborodov(2017),LuKumar(2018),Margalit(2019)}.
Here we consider a simple geometric model to illustrate how NS precession affects the
arrival times of radio bursts from NSs.  We use the terminology ``arrival times'' for FRB emission in our model, but note there are no propagation effects.

Figure~\ref{fig:NSPrecDiag} presents the setup for our emission model, in the frame co-rotating with the NS (body frame), with (effective) principal axis $\hI_i$ defining an orthogonal coordinate system.  The NS spin axis $\hom$ is inclined to $\hI_3$ by an angle $\theta$ ($\cos \theta = \hom \bcdot \hI_3$), and precesses about $\hI_3$ at the period $\Pprec$ (eq.~\ref{eq:Pprec}; see eq.~\ref{eq:hom(t)}).  
An observer views the NS in a direction $\hn$ constant in the inertial frame, but rotating 
about $\hom$ in the body frame with inclination $\nu$ ($\cos \nu = \hn \bcdot \hom$) 
and spin period $P$. Note that in the body frame of the NS, $\hn$ satisfies the equation
${\der \hn}/{\der t} + \bom \btimes \hn = 0$.
Since $\hom$ evolves over a timescale much longer than $\hn$ ($|\der \hom/\der t|/|\der \hn/\der t| \sim \epseff \ll 1$), we can treat $\hom$ as approximately constant to obtain
\begin{align}
    \hn(t) = \ & \frac{\sin\nu \cos\vphi_n}{\sin\theta} (\hI_3 \btimes \hom) \btimes \hom
    \nonumber \\
    &- \frac{\sin\nu \sin\vphi_n}{\sin\theta} (\hI_3 \btimes \hom)  + (\cos\nu)\hom
    \label{eq:hn(t)}
\end{align}
where 
\be
\vphi_n(t) = \om t + \vphi_{n0}
\label{eq:vphi_n}
\ee
is the rotation phase of $\hn$ around $\hom$, with $\vphi_{n0} = \vphi_n(0)$. 
The NS's dipole axis $\hp$ is fixed in the body frame, inclined to $\hI_3$ by an angle $\psi$ ($\cos \psi = \hp \bcdot \hI_3$).  For concreteness, we take $\hp$ to lie in the plane spanned by $\hI_1$ and $\hI_3$.  The inclination between $\hn$ and $\hp$ is specified by the angle $\xi$ ($\cos \xi = \hn \bcdot \hp$).  We consider a phenomenological emission model as an example, where we assume the radiation is emitted from a cone centered at $\hp$ with opening angle $\ag$, with the emission intensity $I$ tapering off as $\hn$ becomes more misaligned with $\hp$:
\be
I = I_0 \exp \left(-\frac{\xi^2}{2 \ag^2} \right).
\label{eq:I}
\ee

Figure~\ref{fig:NSEmit} shows two examples of the FRB emission pattern produced in our model.
Although reproducing the periodicity of FRB 180916 requires $\Pprec = 16.35 \ \days$, 
the NS spin period $P$ and effective ellipticity $\epseff$ remain unconstrained (but related, see eq.~\ref{eq:Pprec}).  To leave $P$ unconstrained, and to add stocaticity to our simple emission model, we evaluate $\hom(t)$ (eq.~\ref{eq:hom(t)}) at $N$ times ($N=400$ for the example), which we denote by $t_i$, spread linearly between $t_1 = 0$ to $t_N = \tobs$.  We then pick $\vphi_n(t_i) = \vphi_i$ (eq.~\ref{eq:vphi_n}) randomly from a uniform distribution over the interval $[0,2\pi]$.  
The observer's orientation $\hn_i = \hn(t_i)$ is then evaluated with equation~\eqref{eq:hn(t)}, and the FRB emission $I$ at time $t = t_i$ is computed with equation~\eqref{eq:I}.

We see from Figure~\ref{fig:NSEmit} 
that, despite the simplicity of our model, it does well in reproducing the spacing of the periodic bursts, as well as the clustering of bursts over the precession phase.  The left panel of Figure~\ref{fig:NSEmit} shows some of the $\pm$2.6 day intervals around multiples of $\Pprec$ (light cyan bands) have no detectable bursts, while other epochs have multiple detectable bursts.  More specifically, when $t/\Pprec \approx {\rm integer}$, some draws at times $t_i$ get no instances of $\hn_i \approx \hp$, while other draws get multiple instances of $\hn_i \approx \hp$, due to the changing phase of $\hn$ around $\hom$.  Notice that no bursts are detected at intervals away from integer multiple of $\Pprec$. This occurs because over most of the spin precession phase $\vphi_\om$, $\hom$ lies far from $\hp$, and $\hn$ closely follows $\hom$.  The right panel of Figure~\ref{fig:NSEmit} shows that the burst intensities $I$ are clustered around the phase 0.5, with a spread which depends on the model parameters (angles $\psi$, $\theta$, \& $\nu$, see Fig.~\ref{fig:NSPrecDiag}).  Over this spread, the burst intensities vary by two orders of magnitude, with little dependence on $\vphi_\om$.  All these features were seen in FRB 180916 \citep{CHIME(2020)}.

Figure~\ref{fig:NSEmitPhase} is the same as Figure~\ref{fig:NSEmit}, except instead of drawing a single value 
of the rotational phase $\vphi_n$ at $t = t_i$, we pick $N_n$ linearly-spaced values spanning the interval $[0,2\pi]$.  With many more points sampled for $\vphi_n$, we see the FRB emission is confined to the light cyan epochs. The intensity profile shape with the FRB phase and
the amount of clustering around phase 0.5 depend on the model parameters ($\psi$, $\theta$, $\nu$).

\section{Polarization}
\label{sec:Discuss}

\begin{figure}
	\includegraphics[width=\columnwidth]{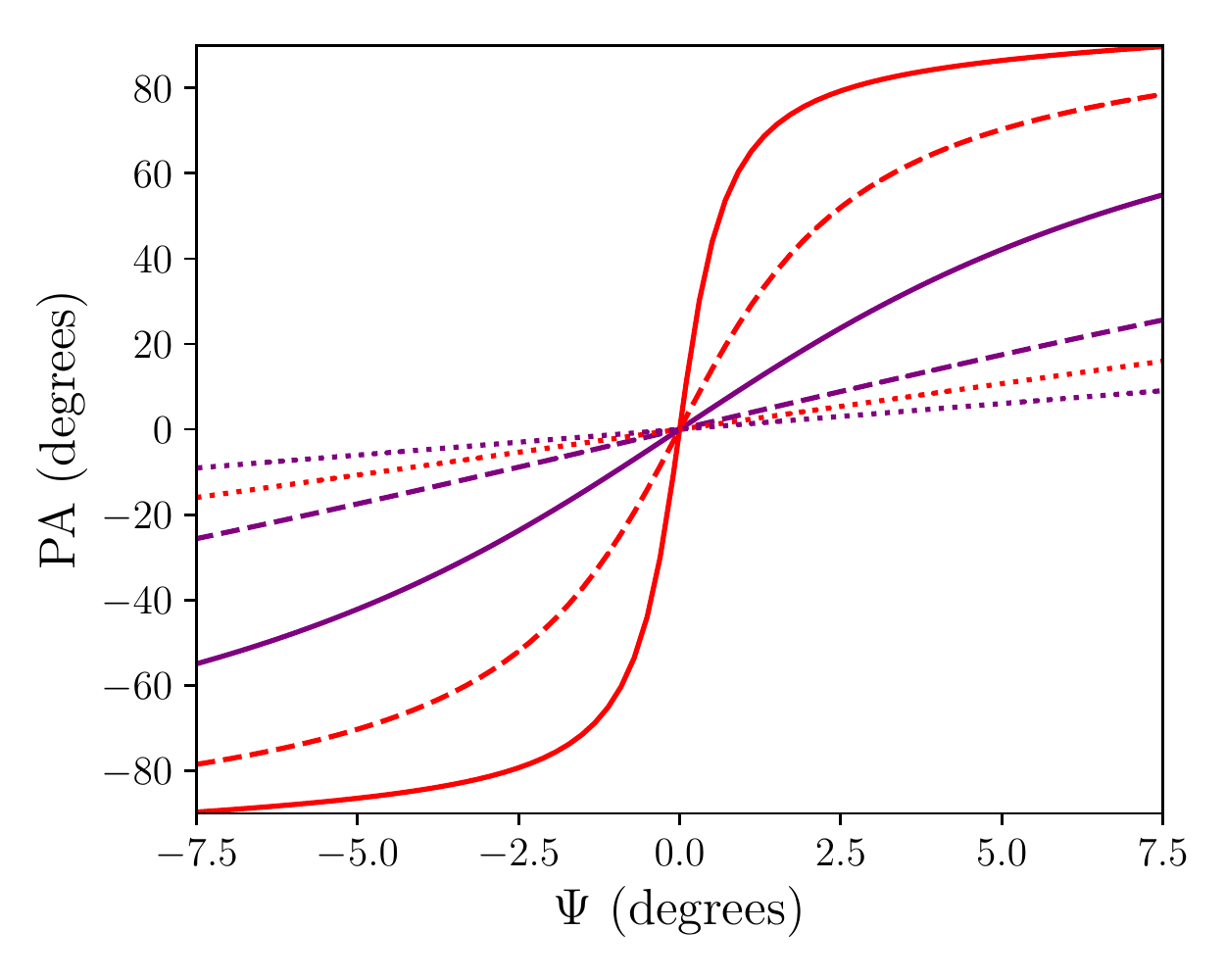}
    \caption{
    Position angle of polarization ${\rm PA}$ with rotational phase $\Psi$ of $\hp$ around $\hom$, for $\vphi_\om = 10^\circ$ (solid), $\vphi_\om = 20^\circ$ (dashed), and $\vphi_\om = 180^\circ$ (dotted).  The model parameters are $\psi = 3^\circ$ (red) and $\psi = 7^\circ$ (purple), with $\theta = 10^\circ$ and $\nu = \theta - \psi$.
    }
    \label{fig:PA}
\end{figure}

In our simple model, the rotational frequency $\om$ and the angles $\theta$, $\psi$, and $\nu$ (see Fig.~\ref{fig:NSPrecDiag}) are constant over timescales much shorter than the spindowm time $\tsd$ (eq.~\ref{eq:tsd}), and thus the FRB emission pattern is constant. However, the magnetic 
obliquity $\chi$ (angle between $\hp$ and $\hom$) is modulated with period $P_{\rm prec}$.
This can change the short-term (on the timescale of rotation period $P$) polarization pattern
of the emission \citep[e.g.][]{Weisberg(2010)}. In particular, if we use the rotating vector model to describe the 
linear polarization from the FRBs \citep[e.g.][]{RadhakrishnanCooke(1969),Wang(2010),Lu(2019)}, the shape of 
polarization sweep (as a function of the NS rotation phase) will be modulated with period 
$P_{\rm prec}$:
\be
{\rm PA} = \tan^{-1} \left( \frac{-\sin \chi \sin \Psi}{\cos \chi \sin \nu - \sin \chi \cos \nu \cos \Psi} \right).
\label{eq:PA}
\ee
Here, ${\rm PA}$ is the polarization position angle (measured from the
projection of $\hom$ in the sky plane), and $\Psi$ is the rotational phase of the NS dipole axis $\hp$ around the rotation axis $\hom$.  Figure~\ref{fig:PA} displays how the polarization angle ${\rm PA}$ is modulated by a precessing NS.  Since we require the line of sight to be almost parallel to the dipole axis to observe FRB emission ($\hn \approx \hp$), a precessing NS can sigificantly affect the ${\rm PA}$ sweep across the rotational phase.
Note that the "mean" polarization position angle (as determined by 
the projection of the rotation axis in the sky plane) is unchanged.

Over timescales comparable to or longer than the spindown time,
$\psi$ remains constant (as we assume $\hp$ is frozen in the NS) and
$\nu$ is also constant to a good precision (since $|\epseff| \ll 1$), but $\om$ and $\theta$ will 
decrease over time (see Fig.~\ref{fig:SpinDown}).
This will lengthen the NS precession period $\Pprec$ (eq.~\ref{eq:Pprec}) and
induce a secular change in the magnetic obliquity $\chi$ (see Fig.~\ref{fig:SpinDown}), which 
in turn will affect the polarization sweep.


\section{Conclusions}
\label{sec:Conc}

We have shown that free precession of isolated neutron stars can in principle
explain the observed 16 day periodicity of FRB 180916. The precession arises either 
from the aspherical deformation of the neutron star by strong internal magnetic fields or from
the "effective" deformation associated with the near-zone dipole or multipole fields coroting 
with the star. The required field strength is of order $10^{15}$~G, depending on the dominant 
deformation mechanism (see Fig.~\ref{fig:P_B_constraints}). Using a simple geometric FRB emission model, where 
radio bursts are emitted along the magnetic dipole axis, we show that the emission pattern 
from a precessing magnetar (Figs.~\ref{fig:NSPrecDiag}-\ref{fig:NSEmitPhase}) can match that observed from FRB 180916.
The fact that a stable precession period has been detected in FRB 180916 during 410 days
of observation implies that the neutron star spin frequency $\omega$ satisfies
$\omega R/c\ll 1$, i.e., the spin period is much larger than milli-seconds, since a small spin period would lead to a rapidly-changing FRB periodicity (Fig.~\ref{fig:SpinDown}).
Our simple model also predicts distinct variations in the polarization profiles for the FRB emission;
these may be tested by future observations.

Needless to say, our simple geometric FRB emission model is highly idealized. Therefore the 
emission pattern and polarization profile presented in this paper are for illustrative purpose
only. But it is likely that any beamed emission that originates from inside a corotating 
magnetosphere will share qualitatively similar characteristics as our simple model.

\section*{Acknowledgements}

We thank the anonymous referee, whose comments improved the clarity of this work.  JZ thanks Dongzi Li, Ue-Li Pen, and Chris Thompson for useful discussions.  This work was supported in part by NSF grant AST-17152.  JZ is supported by a CITA postdoctoral fellowship.

\bibliography{NS_Prec_FRB}{}
\bibliographystyle{aasjournal}

\end{document}